\documentclass[aps,prl,twocolumn,groupedaddress,showkeys]{revtex4}
\usepackage{epsfig}

\begin{document}


\title{A statistical view on team handball results: home advantage, team fitness and prediction of match outcomes}

\author{Jens Smiatek$^1$}
\author{Andreas Heuer$^1$}
\affiliation{$^1$Institute of Physical Chemistry, University of M{\"u}nster, D-48149 M{\"u}nster, Germany}

\begin{abstract}
We analyze the results of the German Team Handball Bundesliga for ten seasons in a model-free statistical time series approach. 
We will show that the home advantage is nearly negligible compared to the total sum of goals. Specific interest has been spent on the time evolution
of the team fitness expressed in terms of the goal difference. In contrast to soccer, our results indicate a decay of the team fitness values over a season while the long time correlation
behavior over years is nearly comparable.
We are able to explain the dominance of a few teams by the  
large value for the total number of goals in a match. A method for the prediction of match winners is presented in good accuracy with the real results. 
We analyze the properties of promoted teams and indicate drastic level changes between the Bundesliga and the second league. Our findings reflect in good agreement recent discussions on modern 
successful attack strategies.
\end{abstract}
\date{\today}
\keywords{Time series analysis of sports results, statistics, interdisciplinary applications of physics}
\maketitle

\section{Introduction}
The interdisciplinary application of statistical physics methods has emerged constantly. 
For example, numerous investigations on time series analysis resulting from successive matches in sports leagues have 
been published in recent years \cite{Heuer1,Heuer2,Heuer3,BenNaim1,BenNaim2}. 
Most of the studies focus on the detailed analysis of soccer characteristics \cite{Heuer1,Heuer2} and the question if the champion is always the best team \cite{BenNaim1,BenNaim2,Skinner}.
By having a closer look, it can be found that often the studies can be separated into phenomenological and model-free approaches.\\
In contrast to the latter studies \cite{Heuer1,Heuer2, BenNaim1,BenNaim2} which often focus on a time series analysis, the phenomenological constructs are based on the estimation of several parameters 
to answer open questions like the prediction of 
match results \cite{Lee,Dixon,Karlis}.\\
In addition to the estimation of match outcomes,
specific interest has been spent on the soccer goal distribution. Recent findings, following a larger discussion \cite{Greenhough,Bittner1,Bittner2} concerning the distribution of goals of a team in a match
have indicated a Poissonian distribution which reveals the interpretation of goals as rare events \cite{Heuer2}.
A few studies also aimed to yield a quantitative expression for the quality of the teams \cite{Heuer1,Heuer2,BenNaim1,BenNaim2}. 
It has been recently validated, that a good variable for a detailed differentiation is given by the average goal difference of a team. It has to be mentioned that in contrast to the public opinion, 
a time-series analysis has indicated  
a nearly constant value over a season \cite{Heuer1}. This fact is also independent of coach changes \cite{Heuer3}
which is often the method of choice for team managers after a long period of failure. 
In addition to the detailed study of soccer characteristics, further results have been also published on the properties of basketball \cite{Kubatko,Gabel} and the analysis of tennis matches \cite{radicchi,OMalley}.\\ 
In contrast to the world-wide interest in soccer, team handball is popular mainly in Europe. A growing interest in the German Handball Bundesliga and in team handball in general 
over the last years can be found which is well reflected by the number of matches broadcasted at TV and the 
increasing number of active non-professional players \cite{dhb}.\\
For this study, we analyzed the specific characteristics of the highest league in german professional team handball.
The German Team Handball Bundesliga consists of 18 teams where  
each team has two matches against the other teams, one match at home and one as a guest. The Bundesliga can be seen as a round-robin tournament without playoffs or a final. 
Thus, a total number of 306 matches divided into 34 match days are played during a season. 
After each season, the teams at the end of the table are 
replaced in the following season by the promoted teams of the second league.\\ 
In this study we focus on a model-free time series analysis of team handball results of the German Bundesliga. The data set consists of ten seasons starting from 2001/2002.
We will show how recently developed methods for the study of soccer characteristics \cite{Heuer1,Heuer2,buch} can be transferred to the analysis of team handball results.  
The direct comparison between team handball and soccer results allows to extract specific characteristics and differences.\\
Our main results involve the time evolution of the team fitness, the specific goal distribution and the home advantage which are compared to 
recently published results for soccer \cite{Heuer1,Heuer2,Heuer3,buch}.  
In addition, we will also analyze the properties of promoted teams and propose an explanation for the success of the best teams. Our results can be used to estimate the stochastic contributions which influence 
a match outcome.
These findings allow to present a simple method for the prediction of match results and match winners in good accuracy to the real outcomes.
In contrast to the public opinion, we will show that the variance of performance differences between handball teams is less pronounced than it has been found for the German soccer league. 
The dominance of a few teams like THW Kiel, SG Flensburg-Handewitt, TBV Lemgo and HSV Hamburg in the last years can be mainly explained by a 
significant increase of the total sum of goals in a match and improved attack qualities \cite{Seco,Spaete}. 

\section{Data basis}
We have downloaded the match results of the German Bundesliga from {\sf http://www.sis-handball.de}. For technical reasons, like a non-constant number of teams in earlier years
we have omitted all seasons before 2001/2002. The last season was 2010/2011. The total number of matches used in this study was 3060.
The second league consists of a northern and a southern division
where each division consists of 18 teams. The best teams are promoted in the following season into the Bundesliga. 
Throughout the manuscript, we denote the values related to the number of goals in a match with $g$ while values over a season are denoted by the capital letter $G$. 
 
\section{Results}
\subsection{Statistics of a match}
\begin{figure}[!h]
\includegraphics[width=0.5\textwidth]{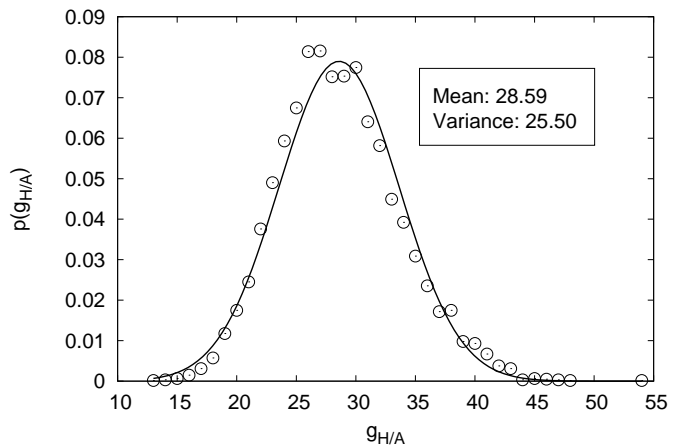}
\caption{Distribution of home goals $g_H$ and guest goals $g_A$.}
\label{fig1}
\end{figure}

\begin{figure}[!h]
\includegraphics[width=0.5\textwidth]{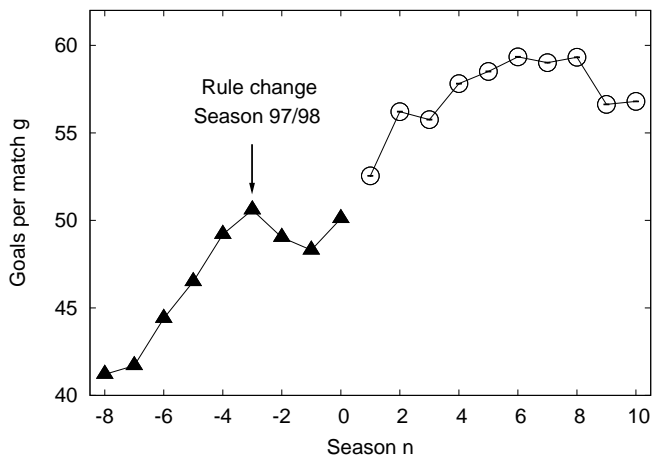}
\caption{The number of goals $g$ per match for a specific season $n$. The circles denote the investigated seasons from 2001/2002-2010/2011. A general increase starting from 1992/1992 from roughly 41 to 59 goals can be observed.}
\label{fig2}
\end{figure}

\begin{figure}[!h]
\includegraphics[width=0.5\textwidth]{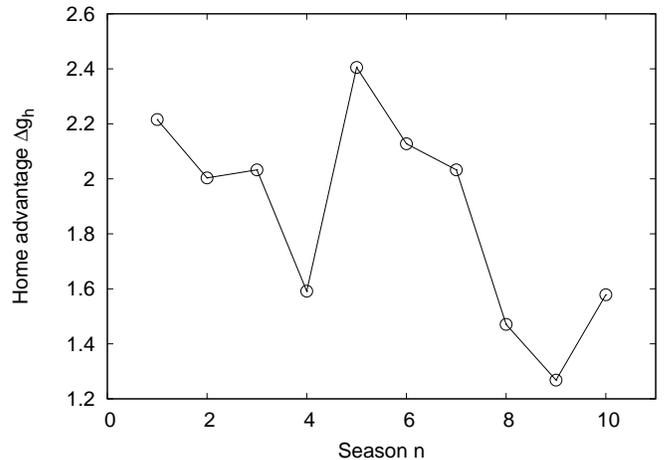}
\caption{Home advantage $\Delta g_h = g_H-g_A$ for all seasons.}
\label{fig12}
\end{figure}

We start our analysis by the investigation of the inherent characteristics of the game. The win probability at home 
is given by 58.9\%. This value coincides well with recent values for the spanish team handball league where a home win probability of 61\% has been found \cite{Aguilar}. 
The guest win probability is given by 33.2\% and a draw between both teams appeared with a probability of 7.9\%.\\ 
For 47.2\% of all teams, a positive goal difference after 34 match days has been found which roughly corresponds to the first eight teams of the table. The percentage of goals scored by these 
teams compared to the total sum of goals is given by 50.8\%. This indicates that eight teams in a season throw more than half of the total number of goals.\\
The largest total number of goals per match has been observed in December 2005 in the match THW Kiel vs.~SC Magdeburg. The result was 54:34. In agreement, the largest number of 
total attacks with 160 was also found in this match \cite{Spaete}. The lowest total number of goals 
was 32 as it was found for the match FA G{\"o}ppingen vs.~SG Willst{\"a}tt/Schutterwald (19:13) in March 2002. The largest goal difference results from the match HSV Hamburg vs.~Wilhelmshavener HV in 
March 2008 (44:17). \\
Some basic properties of a single match like the mean number of goals $g$, the home goals $g_H$ and the away goals $g_A$ as well as the corresponding variances are presented in Tab.~\ref{tab1}.
\begin{table}[!h]
\centering
\caption{Mean number of goals $g$, home goals $g_H$, away goals $g_A$ for a single match and the variances $\sigma^2$ for the corresponding values.}
\label{tab1}
\begin{tabular}{llll}
\hline
& $g$      & $g_H$      & $g_A$ \\
\hline
Mean      & $57.19$      & $29.53$     & $27.66$ \\
$\sigma^2$ & $53.88$      & $25.42$     & $23.81$ \\
\hline
\end{tabular}
\end{table} 
The non-vanishing value for the difference between $g_H$ and $g_a$ indicates the presence of a home advantage which is given by \cite{Heuer1} 
\begin{equation}
\label{eq:adv}
\Delta g_h = g_H-g_A 
\end{equation}
where we have found a value of $\Delta g_h=1.87$.
Comparing this number to the total number of goals gives a ratio of $\Delta g_h/g = 0.033$ which is nearly negligible compared to soccer where this ratio has been determined to $0.17$ \cite{Heuer1,buch}. 
Thus, we conclude as a first point that the home advantage is nearly negligible for team handball by roughly two goals. 
Thus it becomes clear, that long series of home winnings as they have been claimed for SG Flensburg-Handewitt and THW Kiel are related to the quality of the teams and not to an explicit 
home strength.\\
The distribution of home and guest goals $g_{H/A}$ is presented in Fig.~\ref{fig1}. 
It can be seen that the goal distribution can be well described by a Gaussian function with extreme values for $g=13$ and 54. The variance is given by $\sigma^2_{g_{H/A}}=25.5$.
In agreement to recent studies where it has been shown that on average nearly 
50-60 attacks will be performed by a single team in a match \cite{Dreckmann,Brack,Schuessler,Seco}, respectively a total number of roughly 120 attacks for both teams \cite{Brack,Schuessler,Seco}, a goal efficiency of 
approximately 50\% can be estimated. At the end of this section, we will show that these properties are well reflected by Binomial statistics.\\
The significant evolution of the total number of goals in a match as presented in Fig.~\ref{fig2} can be explained by the strict novel interpretation of improved attack strategies like the first and the second wave
as initiated by rule changes in 1997 \cite{Brack,Schuessler,Seco,Secorules}.
It is remarkable that a significant increase from 52 to 59 goals can be indicated from 2001/2002 to 2010/2011 while a general 
increase of roughly 18 goals can be observed from 1992/1993 which is related to emerging professional structures and improved strategies \cite{Secorules}.  
It has to be mentioned that a slight decrease can be also observed for the last two seasons. We assume that this decrease is related to an improvement of defense concepts.\\ 
The detailed values for the seasonal home advantages calculated by Eqn.~\ref{eq:adv} are presented in Fig.~\ref{fig12}.
It can be seen that the home advantage as given by $1.2 - 2.4$ goals is small compared to the total sum of goals. 
In general, smaller values in agreement to soccer \cite{buch} can be observed for the last 3 seasons. 

\subsection{Evolution of team fitness}
\begin{figure}[!h]
\includegraphics[width=0.5\textwidth]{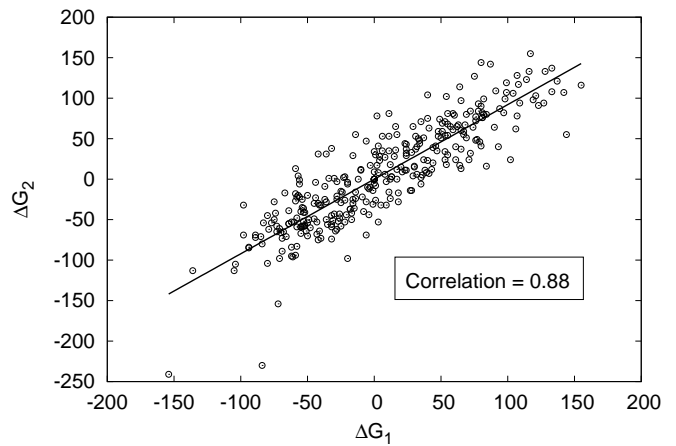}
\caption{The correlation of the goal difference for the first ($\Delta G_1$) and the second half of a season ($\Delta G_2$).}
\label{fig3}
\end{figure}

\begin{figure}[!h]
\includegraphics[width=0.5\textwidth]{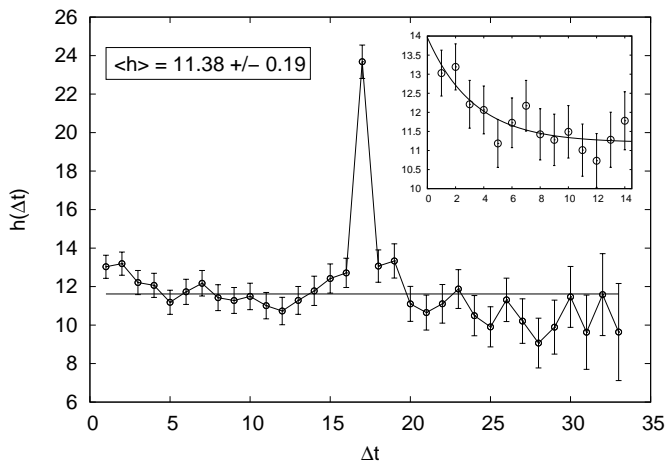}
\caption{Autocorrelation function $h(\Delta t)$ given by Eqn.~\ref{eq:auto} where the straight line represents an average value of $<h>=11.38\pm 0.19$. {\em Inset:} The autocorrelation of 14 successive matches
fitted by Eqn.~\ref{eq:fit}. The corresponding values are $c_1 = 11.2 \pm 0.26$, $c_2 = 2.76 \pm 0.63$ and $\tau = 3.31 \pm 1.58$.} 
\label{fig4}
\end{figure}

\begin{figure}[!h]
\includegraphics[width=0.5\textwidth]{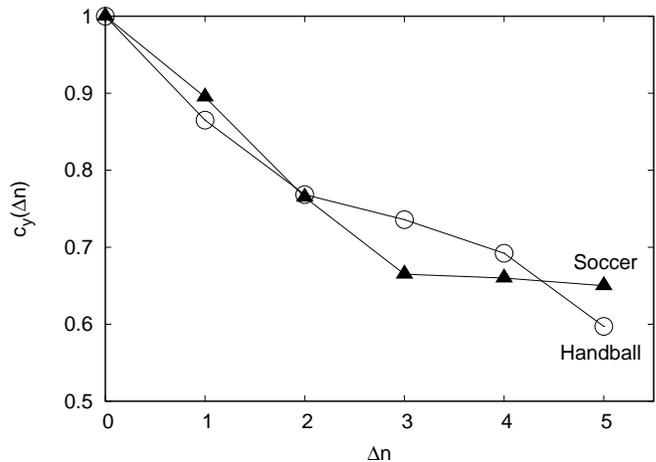}
\caption{Autocorrelation function for the summed goal difference between following half seasons. 
The circles denote the results for handball while the triangles are soccer results from the German Bundesliga (1995/1996-2010/2011) 
which have been published in Ref.~\cite{buch}.} 
\label{fig5}
\end{figure}

We further studied the evolution of the team fitness over a season and between several years. 
The general concept of team fitness allows 
a quantitative description of the performance of the team compared to the other teams \cite{Heuer1,buch}. It has been in detail discussed in \cite{Heuer1}, that the best estimator for the fitness of a soccer team 
is given by the average goal difference of the team per match. 
The underlying studies have also shown that this quantity remains in principle constant for each team over a season with small fluctuations \cite{Heuer1}.
To study this property for handball teams, we 
have plotted the explicit goal difference for each team after a half season (N=17 matches) $\Delta G_1$ and the subsequent half season
$\Delta G_2$ in Fig.~\ref{fig3}.
The correlation between both quantities can be estimated to $r^2=0.88$. 
This large value indicates that the fitness of a handball team is highly correlated over a season in agreement to soccer \cite{Heuer1}.\\
To investigate this point in more detail, we have calculated 
the autocorrelation function 
\begin{equation}
\label{eq:auto}
h(\Delta t)=<\Delta g_{ij}(t_0)\Delta g_{ik}(t_0+\Delta t)> 
\end{equation}
in Fig.~\ref{fig4} as described in \cite{Heuer2}
where $\Delta g_{ij}$ denotes the difference of a match of team i {\em{vs.}} team j with the result $g_i:g_j$. We have found an average value of $<h>=11.38\pm 0.19$ which is represented by the straight line. 
The significant increase at $\Delta t=17$ corresponds to the fact that the opponent team is identical to $t_0$. A further explanation and more insights into the occurrence of this large correlation 
can be found in \cite{Heuer2,buch}. 
A slight non-monotonic decrease of the autocorrelation function can be observed for all match days with $\Delta t\leq 14$. 
To investigate this decay in more detail,
we have fitted the corresponding values with the following function
\begin{equation}
\label{eq:fit}
h(\Delta t) = c_1 + c_2\exp(-\Delta t/\tau)
\end{equation}
which gives  $c_1 = 11.2 \pm 0.26$, $c_2 = 2.76 \pm 0.63$ and $\tau = 3.31 \pm 1.58$. Hence one can conclude, that the decorrelation time of $\tau$ of $3-4$ match days represents a short time decay. 
Therefore one can assume that the outcome of the actual match influences the results that will be obtained in the subsequent 3 to 4 matches.\\
The fluctuating values for $\Delta t > 20$ can
be also observed in the German soccer league \cite{Heuer2}. 
In addition, we are able to observe a small decrease of $h(\Delta t)$ over the complete season in contrast to soccer results \cite{buch,Heuer1}.
Therefore fluctuating values for the team fitness can be identified on short as well as on longer timescales.\\
The long time correlation over years for team $i$ by the goal differences can be studied by the summed values for match days $t=1-17$ given by $\Delta G_{i,1}$ and $t=18-34$ ($\Delta G_{i,2}$). 
Therefore we divided each season into two parts to improve the statistics which results in 
$2n$ values for each team staying for $n$ subsequent seasons in the Bundesliga.  
An expression for the seasonal autocorrelation of the team fitness is given by the relation 
\begin{equation}
c_y(\Delta n) = \frac{<\Delta G_{i,j}(n_0)\Delta G_{i,m}(n_0+\Delta n)>}{<\Delta G_{i,1}(n_0)\Delta G_{i,2}(n_0)>}
\end{equation}
with $j,m=1,2$ denoting the first or the second part for the corresponding season at $n_0$ and $n_0+\Delta n$.  
The autocorrelation value for $\Delta n=0$ corresponds to the correlation of the first and the second half of each single season at $n_0$ which also expresses the normalization constant.\\ 
The results as well as the corresponding values for soccer are presented in Fig.~\ref{fig5}. \\
It is remarkable, that a nearly identical decorrelation time compared to the results derived for soccer in Ref.~\cite{buch} can be observed. 
To explain these effects it was discussed \cite{buch} that short time effects between subsequent years disappear after two seasons. 
We assume that the slightly larger values for team handball as evident for season differences of $\Delta n = 3,4$ can be related to the long time dominance of a few teams.
We therefore propose that economic reasons as discussed for soccer \cite{Heuer1,buch} are also able to explain the decay of the fitness values between subsequent seasons. 
\subsection{Variance analysis: Stochastic contributions}
\begin{figure}[!h]
\includegraphics[width=0.5\textwidth]{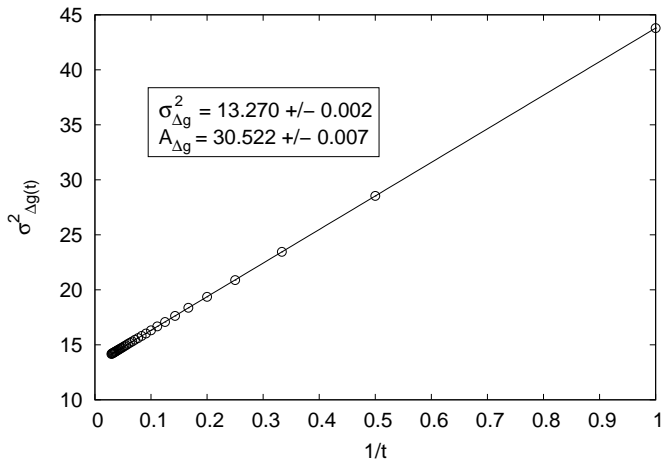}
\caption{The variance of $\Delta g(t)$ averaged over all years and teams. The straight line is a fit according to Eqn.~\ref{eq:var}.}
\label{fig6}
\end{figure}

\begin{figure}[!h]
\includegraphics[width=0.5\textwidth]{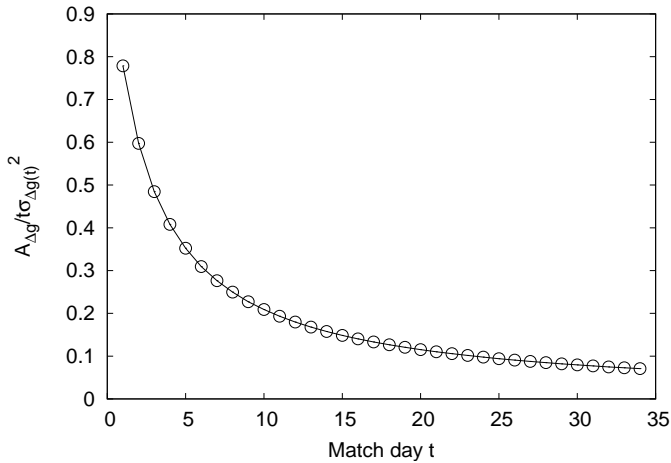}
\caption{Relative stochastic influence on the overall variance of $\sigma_{\Delta g}^2$ after $t$ matches.}
\label{fig7}
\end{figure}

\begin{figure}[!h]
\includegraphics[width=0.5\textwidth]{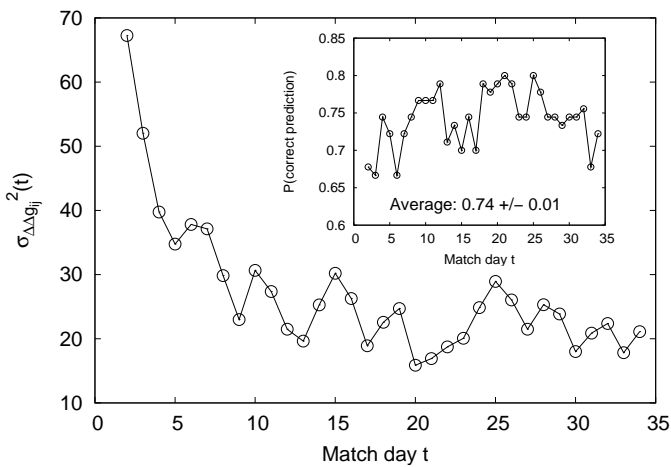}
\caption{Variances of the predicted average match goal differences $\Delta \Delta g_{ij,t}$ to the real results for all match days. 
{\em Inset:} The probability for a correct prediction of a winner for all match days averaged over all seasons.}
\label{fig13}
\end{figure}

It has been discussed in \cite{Heuer1,buch} that stochastic contributions play a dominant role for the outcome of soccer match results. 
A method to determine the strength of these contributions is given by an analysis of the 
corresponding variances. The evolution of the variance for the goal differences for single matches of a team 
can be expressed by
\begin{equation}
\label{eq:var}
\sigma_{\Delta g(t)}^2=\sigma_{\Delta g}^2+\frac{A_{\Delta g}}{t}
\end{equation}
where the stochastic contributions are contained in the second term of the right hand side. The variance for $t\rightarrow\infty$ matches given by 
$\lim_{t\rightarrow \infty}\sigma_{\Delta g(t)}^2 =\sigma_{\Delta g}^2$ estimates
the variance of the team fitness in the limit of negligible stochastic contributions and
can be obtained by a linear regression.
The term $A_{\Delta g}/t$ contains the stochastic contributions for the total of $t$ matches played. The method and the derivation of Eqn.~\ref{eq:var} is in detail explained in Refs.~\cite{Heuer1,buch}. 
To improve the statistics we have not only used the first $t$ matches 
of a season but used all sets of $t$ successive matches of a team for the averaging. In addition, we also subtracted the home advantage for the specific season from all results which gives the neutral 
goal difference $\Delta g = (g_H-1/2\Delta g_h)-(g_A+1/2\Delta g_h)= g_+ - g_-$.
The values are shown in Fig.~\ref{fig6}. 
Regarding the linear regression, we have found $\sigma_{\Delta g}^2=13.27\pm 0.0002$ and $A_{\Delta g}=30.522\pm 0.007$. We also performed an identical analysis for the 
goals $g_+$ and the goals against $g_-$ as defined above. The corresponding values for team handball and soccer as derived in \cite{Heuer1,buch} are presented in Tab.~\ref{tab3}.\\
\begin{table}[!h]
\centering
\caption{Variances $\sigma^2$ for $t\rightarrow\infty$ matches and stochastic term $A$ according to Eqn.~\ref{eq:var} for the neutral goal difference $\Delta g=g_+-g_-$, goals $g_+$ and goals against $g_-$ 
for team handball results denoted by the subscript $HB$ and soccer results given by the subscript $S$.}
\label{tab3}
\begin{tabular}{llll}
\hline
Values & $\Delta g$      & $g_+$      & $g_-$ \\
\hline
$\sigma_{HB}^2$      & $13.3$      & $8.2$     & $3.7$ \\
$A_{HB}$      & $31$  & $17$     & $22$ \\
$\sigma_S^2$      & $0.24$      & $0.076$     & $0.06$ \\
$A_S$      & $3.0$  & $1.7$     & $1.3$ \\
\hline
\end{tabular}
\end{table}
It has to be mentioned that a direct comparison between the values for both team sports is not possible due to the different total numbers of goals per 
match.
Therefore the ratio of the total number of handball goals and soccer goals in a match has to be determined. 
This value is given by $g_{HB}/g_S = 57.19/2.75\approx 20.8$ as mentioned in \cite{buch}.\\ 
The transferred soccer terms for an average number of $g=57.19$ by regarding Tab.~\ref{tab3} are given by $A_{\Delta g,S} = A_S\cdot 20.8$ and $\sigma_{\Delta g,S}^2 = \sigma_S^2\cdot(20.8)^2$ 
which gives for the ratio
$A_{\Delta g,S}/\sigma_{\Delta g,S}^2=0.64$. 
The identical relation for team handball yields $A_{\Delta g, HB}/\sigma_{\Delta g, HB}^2=2.3$. The comparison between both ratios indicates the influence of the 
stochastic contributions on the team fitness \cite{Heuer1,buch}. 
Thus, it can be clearly seen that if the typical sum of goals in a soccer match would be identical to team handball, a general smaller stochastic contribution can be found for soccer results.
This further implies that the differences in the fitness for different teams in soccer are broader compared to handball and vice versa \cite{buch}. 
Hence, we can conclude that the seemingly significant team fitness differences present in team handball are mainly related 
to the large number of total goals and not directly to the distinct qualities of the teams.\\
To study the characteristics of the handball goal distribution of a team per match, it is worthwhile to compare the ratio $A_{\Delta g}/<g>$ to the goal efficiency of a team as it has been discussed in Ref.~\cite{Heuer2}
and previously in this section. 
Inserting the values of Tab.~\ref{tab3}, we have found a ratio of $A_{\Delta g}/<g>=0.55$ while the goal efficiency is roughly $\approx 0.5$ as mentioned previously in this section. 
Hence it can be concluded that throwing a goal is comparable to throwing a coin due to a nearly identical success probability. Therefore the Poissonian statistics which have been found for soccer \cite{Heuer2}
have to be replaced by Binomial statistics for team handball. We finally conclude that the distribution of goals per team in a match is given by a Binomial distribution function which is also validated 
by the ratio $A_{\Delta g}/<g>=0.55$ in good agreement to the expected value of 0.5.\\
To estimate the influence of stochastic contributions on a match outcome, we have calculated the values for the relation $(A_{\Delta g}/t)/\sigma_{\Delta g(t)}^2$ according to Eqn.~\ref{eq:var} as presented 
in Fig.~\ref{fig7}.
Compared to soccer\cite{Heuer1}, where it was found that the stochastic contributions decay to $1/e$ after 30 match days, we have found an identical decay for team handball after 7 match days. 
This
observation mainly relies on the larger total number of goals as it has been discussed above. One can conclude, that this fact in combination with the corresponding smaller stochastic contributions leads to 
a meaningful table after the tenth match day which displays the real quality of the teams in good accuracy.
It becomes obvious that the variances 
for the average goal difference which express the team fitness follow a significant evolution in time with a steady decrease of the stochastic contributions.\\
The smaller stochastic contributions to a single match compared to soccer also allow a better prediction of match results.
To investigate this in more detail,
we have taken the average goal difference $\Delta g_{i,t}(n_0)$ of a team at the $(t-1)$-th match day of a season $n_0$ and have estimated the match goal difference at match day $t$ 
between two teams $i,j$ by 
$\Delta\Delta g_{ij,t} = \Delta g_{i,t-1}(n_0)-\Delta g_{j,t-1}(n_0)$ for the outstanding match at the $t$-th match day. 
By adding the home advantage, we are able to compare our predictions with the real results. 
The corresponding variances for a single match resulting from comparison with the real results are shown in Fig.~\ref{fig13}.
It can be clearly seen that a significant decrease can be observed until the tenth match day in agreement to the results shown in Fig.~\ref{fig7} which indicates better predictions.\\ 
Furthermore we were also able to estimate the winning 
team by the same procedure. The probability of a correct prediction is given by an accuracy of 74\% as shown in the inset of Fig.~\ref{fig13}. It has to be mentioned that this value is significantly larger than for soccer,
where the correct winner prediction is around 52\% \cite{buch}. 
\subsection{Attack and defense strategies and the properties of promoted teams}
\begin{figure}[h!]
\includegraphics[width=0.5\textwidth]{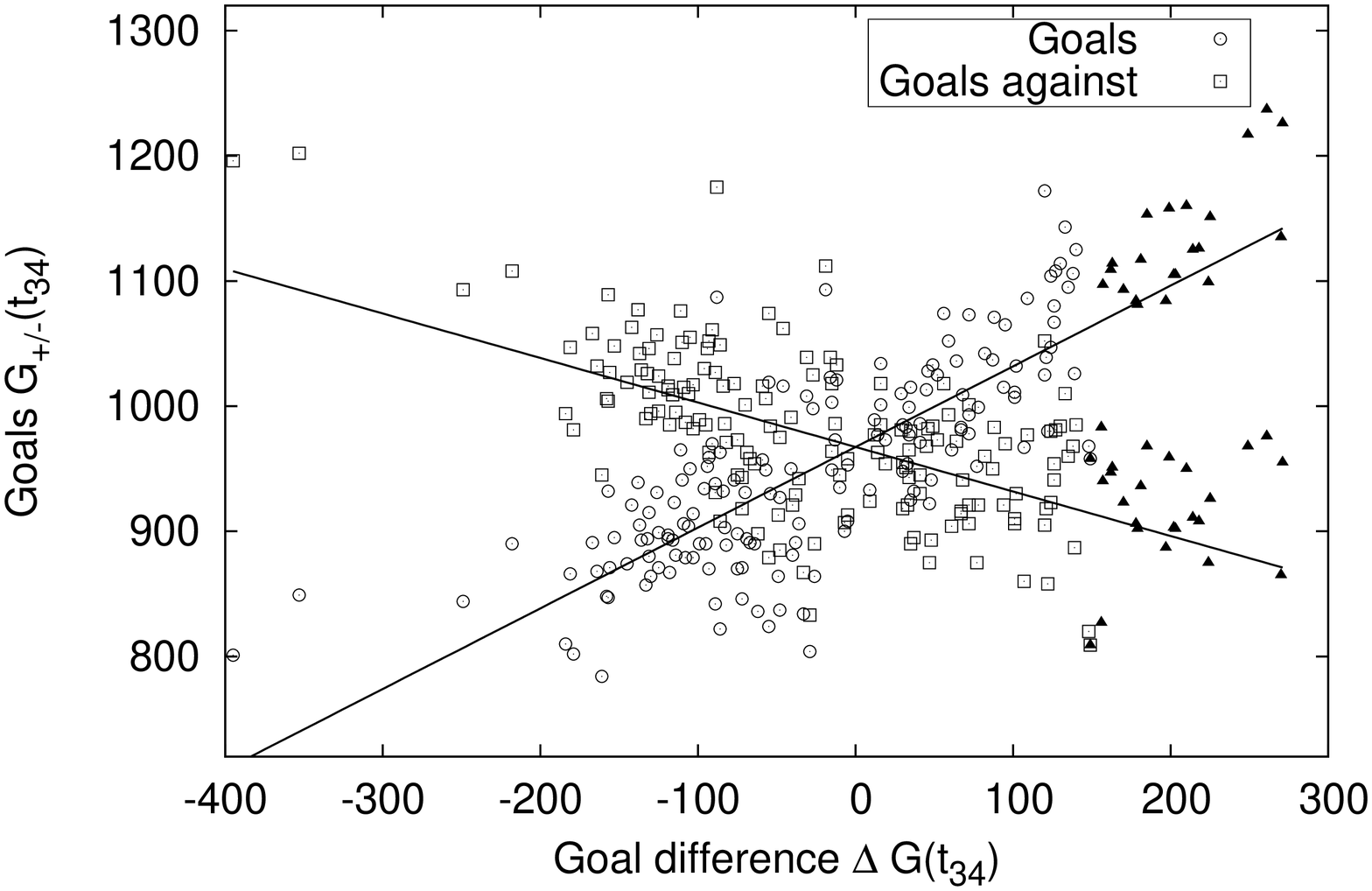}
\caption{Goal difference after 34 match days and the corresponding goals $G_+$ and goals against $G_-$. Teams with goal differences larger than 150 are plotted as filled triangles.}
\label{fig8}
\end{figure}

\begin{figure}[!h]
\includegraphics[width=0.5\textwidth]{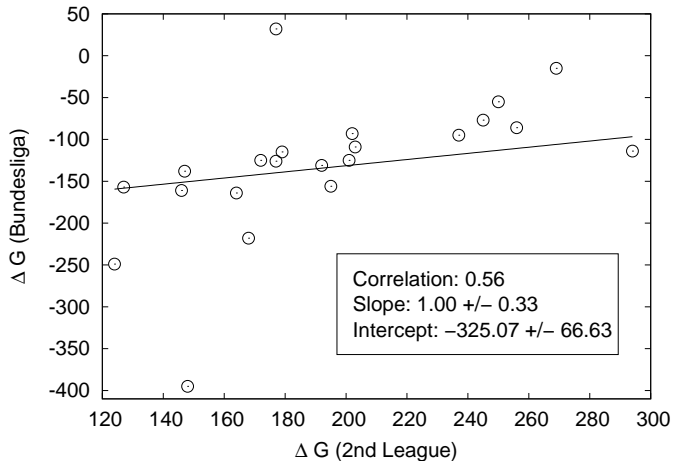}
\caption{Values of the goal difference for the league climbers in the Bundesliga and the second league.}
\label{fig10}
\end{figure}

It has been recently discussed that the strategy of successful handball teams focuses more on the attack instead of the defense \cite{Spaete,Seco,Brack}. 
This assumption is validated by Fig.~\ref{fig8}, where the goal difference for each team after one season with 34 match days is plotted
against the scored goals $G_+$ and the goals against $G_-$. 
The teams which have a goal difference $\Delta G(t_{34})$ larger than 150 are printed as filled triangles. By linear regression, we have calculated the slopes for the scored goals 
and the goals against for all teams with a goal difference $\Delta G \leq 150$.   
We have found values of 0.64 for the attack and $-0.36$ for the defense. 
It can be seen that the results for the scored goals and the goals against for teams with $\Delta G \geq 150$ are significantly located above the corresponding slopes. 
Thus we conclude, that the best teams over all years have a stronger attack compared to the rest of the league while the defense is slightly more worse in agreement to recent discussions \cite{Spaete,Seco,Brack}.
This fact has been often assumed in the literature but was not analyzed in full detail \cite{Spaete}.\\
The general correlation coefficients for teams with a positive goal difference $\Delta G_{\geq 0}$ and a 
negative goal difference $\Delta G_{< 0}$ between the number of scored goals $G_+$ and goals against $G_-$ and their corresponding goal differences $\Delta G(t_{34})$ after 34 match days are presented in Tab.~\ref{tab5}.\\
\begin{table}[!h]
\centering
\caption{Correlation coefficients between the goal difference $\Delta G(t_{34})$ for teams with a positive $\Delta G_{\geq 0}$ and a negative goal difference $\Delta G_-$ 
with the number of scored goals $G_+$ and goals against $G_-$.}
\label{tab5}
\begin{tabular}{lll}
\hline
& $G_+$      & $G_-$ \\
\hline
$\Delta G_{\geq 0}$ & 0.797 & -0.231\\
$\Delta G_{< 0}$ & 0.449 & -0.597\\
\hline
\end{tabular}
\end{table}
The occurrence of an asymmetric behavior can be clearly identified. Teams with positive goal differences mainly achieve a stronger correlation of $\Delta G(t_{34})$ 
with the attack (0.797) while the correlation with the defense 
is significantly weaker (-0.231). In contrast, we have found for teams with a negative goal difference a larger correlation coefficient of -0.597 with the defense while the attack is weaker 
correlated (0.449).
The direct correlation between attack and defense is given by 0.403 for teams with a positive goal difference and 0.449 for teams with a negative goal difference. These values are in complete disagreement to 
recent results for soccer where a strong correlation between attack and defense has been validated \cite{buch}.\\ 
To investigate these findings in more detail, we will show that the analysis of the variances $\sigma_{g_+}^2, \sigma_{g_-}^2$ for the goals and the goals against in a match as presented in Tab.~\ref{tab3} 
can give further insights.
The ratio $\sigma_{g_+}^2/\sigma_{g_-}^2$ for team handball is given by $2.2$. This indicates that the attack properties expressed by the number of thrown goals are broader distributed than the number of goals against. 
For soccer, the corresponding ratio gives $1.27$ which is significantly smaller. Hence, due to the study of these values as well as the differences for the study of the corresponding slopes in 
Fig.~\ref{fig8}, it can be concluded that the attack qualities for handball teams are significantly broader distributed than the defense qualities in agreement to Fig.~\ref{fig8}.\\ 
As a last point we have calculated the properties of promoted teams. The values for the goal difference $\Delta G$ after 34 match days for season $n$ in the second league and the subsequent $n+1$ season 
in the Bundesliga have been calculated and presented in 
Fig.~\ref{fig10}. It can be clearly seen that all teams have a negative or negligible positive goal difference in the following Bundesliga season. The linear regression of the goal difference follows 
a slope of 1 and the 
intercept is given by $\approx -325$ goals, These values correspond well to the intuitive observation, that the establishment of promoted teams in the Bundesliga is a very challenging task. 
The general large value 
of -325 goals demonstrates the qualitative differences between the Bundesliga and the second league.
\section{Discussion}
We have analyzed the results of the German Handball Bundesliga for 10 seasons starting from 2001/2002. Our findings have shown a significant increase of the sum of goals per match in the last years.
We are able to explain this increase by novel attack strategies which allow a rapid turnover and an acceleration of the game. 
Despite this increase, we have found a nearly constant home advantage represented by roughly two goals. It has to be noticed, that compared to the total sum of goals, the home advantage is nearly negligible.\\
In contrast to soccer, we have found a binomial distribution function for the number of goals scored by a team in a match. A simple picture for the goal efficiency of an attacking team is given
by the throwing of a coin.\\ 
In the last years, the dominance of a few teams in the Bundesliga has lead to a larger discussion. It has been argued, that the disappearance of surprises like the winning of outsiders
would lead to a minor public interest.
In agreement to these arguments, we have clearly shown that the stochastic contributions to a handball match which represents the mentioned surprises are significantly smaller compared to soccer. 
However, by a direct transformation of 
the corresponding values, we were able to show that the intuitive dominance of these teams is exclusively related to the large number of total goals in a match.\\
By analyzing the results for the team fitness expressed by the goal difference, we have indicated a short time decay within 5 match days after the fitness correlation reaches a constant value. 
In general, a slight decrease over the season has been also validated.
These findings are in disagreement to soccer results, where nearly constant fitness values over 34 match days has been indicated.
We assume that these variations can be explained by the
long duration of the season, mostly from September to June. In addition a lot of the players of the 
German Bundesliga are involved in national teams. It has to be noticed that each season is intercepted by a World or an European championship which facilitate 
injuries as well as an exhaust of the players. We propose that these additional matches as well as training effects result in decreasing fitness values. \\
In addition, we have also found a nearly identical decay between subsequent seasons for the long time fitness behavior compared to soccer results. 
These findings give a first hint that economic facts and other circumstances as discussed in Ref.~\cite{buch} are also relevant for team handball.\\
Finally we have shown that successful teams strengthen their attack instead of the defense. 
These findings are in good agreement to recent studies concerning the 
development of successful team strategies \cite{Spaete,Brack,Schuessler,Seco}. In contrast to soccer, 
where the correlation between 
offense and defense is nearly identical, we have validated that the offense in team handball is more pronounced, specifically for the best teams.\\
Due to the small stochastic contributions to the match results after ten match days, we have proposed a simple prediction method for the outcome of handball matches. We have been able to determine 
the winning team with a probability of roughly 74\%.\\
Hence we assume, that the conceptual analysis of team handball and soccer results which has been presented previously and in this work can be also applied to other types of sports, {\em e. g.} basketball and may
help to offer a systematic comparison.
\section{Acknowledgments}
We thank O. Rubner, D. Riedl, B. Strauss and H.-J. Smiatek for helpful and inspiring discussions.   

\end{document}